\documentclass[prd,preprintnumbers,showpacs,12pt]{revtex4}
\usepackage[active]{srcltx}
\usepackage{epsfig}
\usepackage{graphicx}
\usepackage{bm}
\usepackage{amsmath}
\usepackage{amssymb}
\usepackage{color}

\newcommand{\be}{\begin{equation}}
\newcommand{\dd}{\displaystyle}
\newcommand{\ee}{\end{equation}}
\newcommand{\bea}{\begin{eqnarray}}
\newcommand{\eea}{\end{eqnarray}}

\def\th{\theta}

\def\dbth{\dot {\bar{\th}}}
\def\bth{\bar\theta}
\def\s{\sigma}
\def\a{\alpha}
\def\f{\frac}

\def\da{\dot{a}}
\def\done{{\dot 1}}
\def\dtwo{{\dot 2}}
\def\bth{\bar\theta}

\newcommand{\nn}{\nonumber}
\newcommand{\de}{\partial}

 \def\slash#1{\setbox0=\hbox{$#1$}#1\hskip-\wd0\dimen0=5pt\advance
       \dimen0 by-\ht0\advance\dimen0 by\dp0\lower0.5\dimen0\hbox
         to\wd0{\hss\sl/\/\hss}}

\begin{document}

\title{Superconformal interacting  particles}

\author{Roberto Casalbuoni}\email{casalbuoni@fi.infn.it}
\author{Daniele Dominici}\email{dominici@fi.infn.it}
\affiliation{Department of Physics and Astronomy, University of Florence and
INFN, Florence, Italy}
\author{Joaquim Gomis} \email{joaquim.gomis@ub.edu}
\affiliation{$^ {b)}$Departament de F\'isica Qu\`antica i Astrof\'isica \\
and Institut de Ci\`encies del Cosmos (ICCUB), Universitat de Barcelona,  Martí i Franquès 1, 08028 Barcelona,
Spain}

\begin{abstract}
The free massless superparticle is reanalysed, in particular
 by performing the Gupta-Bleuler quantization, using the first and second class constraints of the model, and obtaining,  as a result, the Weyl equation for the spinorial component of the chiral superfield. Then we construct a superconformal model of two interacting massless  superparticles from  the free case
by the
introduction of an invariant interaction. 
The interaction introduces an effective mass for each particle by modifying the structure of fermionic constraints, all becoming second class. The quantization of the model produces a bilocal chiral superfield. We also generalise the model by considering a system of superconformal interacting particles 
 and its continuum limit.

\end{abstract}
\maketitle

\section{Introduction}\label{sec:0}

The application of conformal  invariance  to classical interacting relativistic  particles has recently been studied 
\cite{Casalbuoni:2014ofa,Casalbuoni:2023bbh}. The motivation was to generalise the non-relativistic one-dimensional case, as for example the Calogero-Moser rational model \cite{Calogero:1970nt,Moser:1975qp,Olshanetsky:1983wh}, which describes $N$ interacting particles via two body interactions. This model is very important  in the context of integrable models.
The other example, always in one dimension, is the conformal quantum mechanics
\cite{deAlfaro:1976vlx}. Since there are also supersymmetric extensions of these models \cite{Fubini:1984hf,Ivanov:1988it,Freedman:1990gd}, we generalise the model contained in \cite{Casalbuoni:2014ofa} to a superconformal one. 

In this paper we have reanalysed
the free massless superparticle and its superconformal symmetries \cite{Casalbuoni:1975bj,Casalbuoni:1976tz,Brink:1981nb,Hori:1986mr}. The superinversion is an important tool to study the superconformal special transformations and to build the invariants   \cite{Buchbinder:1998qv,Park:1997bq}.

As it is well known  the massless Lagrangian implies a mixture of first class and second class fermionic constraints \cite{Hori:1986mr,Brink:1987bi}. By using the light cone variables
it is possible  to disentangle the first and second class constraints  in a non-covariant way and then  perform the Gupta-Bleuler quantization of the system;
as a result we obtain the Weyl equation for the spinorial component of the chiral superfield.

Then we construct a superconformal model of two interacting massless  superparticles from  the free case by using the einbein formulation for the action. The construction of the interaction term
heavily uses the properties of the variables under superinversion. The interaction term is invariant under the
 diagonal superconformal group.

The interaction introduces an effective mass modifying the structure of fermionic constraints, all fermionic constraints are second class. The quantization of the model produces a bilocal chiral superfield. 

We also generalise the model by considering a system of superconformal particles with nearest neighbor interaction
and by studying its continuum limit.

The organisation of the paper is as follows: in Section II we first review the classical and 
quantum theory of the superconformal particle, in Section III we propose a superconformal model for two interacting particles, in Section IV we generalise it to a system of particles on a one-dimensional  lattice and we study its limit when the lattice spacing is sent to zero. In section V we give an outlook.

\section{ A Superconformal  relativistic particle}\label{sec:1}
In this Section we study the Lagrangian and the Hamiltonian formulation of a single superconformal relativistic particle \cite{Casalbuoni:1975bj,Casalbuoni:1976tz,Brink:1981nb,Hori:1986mr}, by analysing the superconformal symmetries, the structure of the constraints and the Gupta-Bleuler quantization of the model. In particular we will show the appearance of the Weyl equation for the spinorial component of the chiral superfield.

 The superconformal invariant action for a massless relativistic particle is given by 
 \be\label{actionconf}
 S=\int d\tau L=\int d\tau \frac{1}{ 2e}\dot \omega^\mu \omega_\mu.\ee 
 where
 \be
 \dot \omega^\mu=\frac{d \omega^\mu}{d\tau}= \dot x^\mu+i\theta \s^\mu\dbth-i \dot\th\s^\mu\bth,
 \label{eq:1}
 \ee
 and $e$ is a Lagrange multiplier. We suppose to be in a $D=4$ space-time with a flat metric  $g^{\mu\nu}=(-,+,+,+)$ and we  follow the spinor notations of the book of Wess and Bagger \cite{Wess:1992cp}. In particular $
(\sigma^\mu)_{\alpha \dot\alpha}=(-1,\sigma^i),~~~(\tilde\sigma^\mu)^{\dot\alpha\alpha}=\epsilon^{\dot\alpha\dot\beta}\epsilon^{\alpha\beta}\sigma^\mu_{\beta\dot\beta}=(-1,-\sigma^i), i=1,2,3$.

The   Lagrangian $L$ is invariant under the following supersymmetry (SUSY) transformations
\bea
&\delta\theta=\epsilon,~~~\delta\bar\theta =\bar\epsilon,&\nn\\
&\delta x^\mu=i\theta\sigma^\mu\bar\epsilon-i\epsilon\sigma^\mu\bar\theta,~~~\delta e=0,&\eea
 where $\epsilon$ and $\bar\epsilon$ are the SUSY  parameters.

As in the case of conformal invariance, where invariance under Poincar\'e, dilatations and inversion is sufficient to ensure invariance under all the conformal group, also in the case of the superconformal invariance, Poincar\'e, dilatations,  
chiral 
\be x\to x,~~~\theta\to e^{-i\Delta/2}\theta,~~~\bar\theta\to e^{i\Delta/2}\bar\theta,
\ee
SUSY transformations and superinversion are enough to guarantee the invariance under  all the superconformal group  \cite{Park:1997bq,Buchbinder:1998qv}. Therefore, in our case, we need only 
to show  that $L$ is invariant under superinversion.
The superinversion acts upon $\dot\omega^\mu$ as follows
\be
\dot\omega^\mu\to A(x)^\mu{}_\nu\dot\omega^\nu,
\label{eq:4}
\ee
where \cite{Park:1997bq} 
\be
A(x)^\mu{}_\nu=\frac 1 {x^2+\th^2\bar\th^2}\left ( \frac {x^2-\th^2\bar\th^2} {x^2+\th^2\bar\th^2}g^\mu_{\,\,\nu}-2 \frac {x^\mu x_\nu}{x^2+\th^2\bar\th^2}
+2 \epsilon^{\mu}_{\,\,\nu\lambda\rho}\frac {\theta\sigma^\lambda\bar\theta x^\rho}{x^2}\right ).
\ee
The matrix $A(x)$ defining the superinversion satisfies the relation
\be A(x)^T g A(x)=\Omega^2(x)\eta.
\label{eq:20}\ee
By using this property we find
\be
\dot\omega^2\to\Omega^2(x)\dot\omega^2,\label{eq:16}\ee
where
\be
\Omega(x)=\frac 1 {x^2+\theta^2\bar\theta^2}\label{eq:17b}\ee
 (for details see  see (\cite{Park:1997bq}).  The Lagrangian $L$, in eq.~(\ref{actionconf}), is superconformal invariant, assuming the following transformation of the einbein under superinversion
\be
e\to \Omega^2(x)e.\label{eq:18}
\ee

By evaluating the momenta from the Lagrangian  we get
 \be
 p^\mu=\frac{\de L}{\de \dot x_{\mu}}=\f  1 e \dot \omega^\mu ,\quad
 \Pi_e=\frac{\de L}{\de \dot e}=0 ,\label{eq:3.13}
 \ee
 \be
 \Pi_\a=\frac{\de L}{\de \dot \th^\a}
 =-ip^\mu\s_{\mu\a\da} 
 \bth^{\dot\a} ,
 \quad
 \bar\Pi_{\da}=\frac{\de L}{\de \dbth^\a}
 =-i\th^\a p^\mu\s_{\mu\a\da}.
 \ee

We therefore obtain the constraints
 \be
  \Pi_e=0,
 \ee
 and
 \be
 D_\a=\Pi_\a+ip^\mu (\s_\mu \bth)_\a=0,\quad \bar D_{\da} =\bar\Pi_{\da}+ip^\mu (\th\s_\mu )_{\da}=0.\quad
 \label{eq:9}
 \ee
  
 The canonical Hamiltonian 
 is
 \be
H_c=\Pi_e \dot e+\f 1 2 e p^2.
 \label{eq:17}
 \ee
 
 The canonical Poisson brackets for boson and fermion variables are given by
 \be\{x^\mu, p^\nu\}=g^{\mu\nu},\quad
\{\Pi_\a,\th^\beta\}=-\delta^\beta_\a ,\quad \{\Pi_{\da},\th^{\dot\beta}\}=-\delta^{\dot \beta}_{\da},
\label{pb}
\ee
The stability of the primary constraints gives the secondary constraint
 \be
 p^2=0,
 \ee
 which is the mass-shell condition for a massless particle.

%
%
%
%

\subsection{Analysis of the constraints and quantization}
Let us now
analyse the structure of the fermionic constraints, in particular their first and second class character.
 The Poisson brackets of $D_\a$ and $\bar D_{\da}$ are given by
 \be
\{D_\a,D_{\beta}\}=0=\{\bar D_{\da},\bar D_{\dot\beta}\}, \quad
\{D_\a,\bar D_{\da}\}=-2i p^\mu\s_{\mu\a\da},
\ee
with
\be
p^\mu\s_{\mu\a\da}=
\begin{pmatrix}
\sqrt{2}p^+&p^{12}\\ (p^{12})^*&\sqrt{2}p^-
\end{pmatrix},
\label{eq:20a}
\ee
and where we have defined
\be
p^\pm=\f 1 {\sqrt{2}}(p^0\pm p^3),\quad p^{12}=p^1-ip^2.
\ee
and $*$ indicates complex conjugation.
The matrix (\ref{eq:20a}) has zero determinant and its  rank  is two on the surface of the constraint $p^2=0$. 
Therefore we have two first class and two second class constraints. The first class constraints are
\be
\hat D_1=-D_1+\f {p^{12}}{\sqrt{2}p^-}{ D}_{2}=
-\Pi_1+\f {p^{12}}{\sqrt{2}p^-}\Pi_2,\quad \hat {\bar D}_{\dot 1}=-\bar D_{\dot 1}+\f {p^{12}}{\sqrt{2}p^-}\bar  D_{\dot 2}=-\bar \Pi_\done+\f {(p^{12})^*}{\sqrt{2}p^-}\bar \Pi_\dtwo,
\label{eq:37}
\ee
we have
\be
\{\hat D_1,\hat D_1\}=0=\{\hat {\bar D}_{\done},\hat{\bar D}_{\done}\},
\ee
\be
\{\hat D_1,\hat {\bar D}_\done\}=-2\sqrt{2} i p^+-i\sqrt{2}\f {|p^{12}|^2}{p^-}+i\sqrt{2}\f {|p^{12}|^2}{p^-}+i\sqrt{2}\f {|p^{12}|^2}{p^-}=0,
\ee
where use has been made of bosonic first class constraint
$
2p^+p^-=|p^{12}|^2
$.
The second class constraints are 
\be
D_2=\Pi_2+i[(p^{12})^*\bth_\done +\sqrt{2}p^-\bth_\dtwo],\quad
\bar D_\dtwo=\bar\Pi_\dtwo+i[p^{12}\th_1 +\sqrt{2}p^-\th_2],\quad
\ee
since
\be
\{D_2,\bar D_\dtwo\}=-2i \sqrt{2} p^-.
\ee
The extended Hamiltonian includes the first class constraints $\Pi_e$, $p^2$,$\hat D_1$  and ${\hat{\bar D}}_\done$
\be
H_E=\Pi_e \dot e+\f 1 2 e p^2+\mu_1 \hat D_1+\bar\mu_\done {\hat{\bar D}}_\done
\ee
where $\mu_1$ and $\bar\mu_\done$ are arbitrary Grassmann multipliers. 

Notice that there are two Grassmann constraints of first class.  This corresponds to the invariance of our  action 
under an additional local symmetry, the kappa symmetry (see \cite{Siegel:1983hh}). It is given by  ($k$ is an arbitrary time dependent two-component anti-commuting spinor)
\be
\delta x^\mu =\frac i 2\bar\theta\tilde\sigma^\mu\sigma\cdot p\bar k-\frac i 2 k\sigma\cdot p\tilde\sigma^\mu\theta,~~~\delta\theta=\frac 12\bar k\tilde\sigma\cdot p,~~~\delta\bar\theta=\frac 12\tilde\sigma\cdot p k,~~~\delta e =2 i k\dot\theta-2i \dot{\bar\theta}\,\bar k,~~~\delta p^\mu=0.\label{eq:48}\ee
where $p^\mu$ is given in eq.~(\ref{eq:3.13}).

The invariance of the model under the kappa symmetry shows that only half of the Grassmann variables are physical.

At this point one of the possibilities to develop the quantum mechanics of the model is the standard procedure that consists in computing the Dirac brackets and quantising with them. However commutators of canonical operators are in general modified by the presence of second class constraints by making cumbersome the quantization.

Instead of using Dirac brackets we can 
 do the weak quantization  
by using standard commutation relations between canonical operators 
\be
[x^\mu,p^\nu]=ig^{\mu\nu},\quad
[\Pi_\a,\th^\beta]=-i\delta^\beta_\a ,\quad [\Pi_{\da},\th^{\dot\beta}]=-i\delta^{\dot \beta}_{\da},
\ee
and by imposing the first class constraints as the operatorial conditions
\be
p^2 |\Phi>=0,
\ee
\be\hat D_1 |\Phi>=0, \quad {\hat{\bar D}}_\done |\Phi>=0.
\label{eq:46}
\ee
For the second class constraints we use the   Gupta Bleuler procedure in the following way
\be
\quad  {\bar D}_\dtwo|\Phi>=0,\quad  <\Phi |  D_2=0
\label{eq:47}
\ee
with $p^\mu=-i \de/\de x_\mu$, $\pi_i=-i\de/\de \th_i$ and $\bar \pi_i=-i\de/\de \bar\th_i$.
By using eqs.~(\ref{eq:37}) and (\ref{eq:47}), we have
\be
\hat {\bar D}_\done |\Phi>=0 \rightarrow {\bar D_\done}|\Phi>=0.
\ee
So $<\th_\a,\bar\th_{\dot\a},x|\Phi>$ is a chiral superfield $\Phi=\Phi (\th,y)$ where $y^\mu=x^\mu+i \th\s^\mu\bar\th$:
\be
\Phi (\th,y)=\phi (y)+\sqrt{2}\th \psi(y) +\th^2 F(y).
\ee
We still have to impose the first condition of (\ref{eq:46});  let us first change the basis from $\th,\bar\th, x$ to $\th,\bar\th ,y$:
\be
\f {\de}{\de \th^\a} =\f {\de}{\de \th^\a} + i (\s^\mu\bar \th)_\a \f {\de}{\de y^\mu},\quad
\f {\de}{\de \bar\th^{\dot \a}} =\f {\de}{\de \bar\th^{\dot \a}} -i  (\th \s^\mu)_{\dot \a} \f {\de}{\de y^\mu},
\ee
\be
\f {\de}{\de x^\mu}=\f {\de}{\de y^\mu}.
\ee
Therefore
\be
\f {\de}{\de \th^1} =\f {\de}{\de \th^1}-i (\sqrt{2}p^+\bar\th_\done+p^{12}\bar\th_\dtwo),
\quad
\f {\de}{\de \th^2} =\f {\de}{\de \th^2}-i ((p^{12})^*\bar\th_\done+\sqrt{2}p^{-}\bar\th_\dtwo).
\ee

$\hat D_1$ can be written as 
\be
\hat D_1 =-i \left ( -\f \de {\de \th_1}+\f {p^{12}}{\sqrt{2}p^-} \f \de {\de \th_2} \right ).
\ee
We have therefore
\be
0=\hat D_1 \Phi (\th,y)=-i \left ( -\f \de {\de \th_1}+\f {p^{12}}{\sqrt{2}p^-} \f \de {\de \th_2} \right )\Phi (\th,y),
\ee
 which implies for the superfield components the equations of motion
\be
-\psi_1(y)+ \f {p^{12}}{\sqrt{2}p^-} \psi_2(y)=0,
\label{eq:51c}
\ee
\be
F(y)=0,
\ee
Eq.~(\ref{eq:51c}) can be rewritten as the Weyl equation
\be
p^\mu{\tilde\s}_\mu \psi(y)=0.
\label{eq:53}
\ee
using
\be
p^\mu{\tilde\s}_\mu
=\begin{pmatrix}
\sqrt{2}p^-&-p^{12}\\ -(p^{12})^*&\sqrt{2}p^+
\end{pmatrix}
\ee
and in eq.~({\ref{eq:53}}) $p^\mu=-i\de/\de x_\mu=-i\de/\de y_\mu$.
\section{Two massless interacting super-particles}
\label{twosuperconformal}
In order to construct the model, let us first consider  the case of 
two free massless particles:
\be
L_2=\frac 1{2e_1}(\dot\omega_1^\mu)^2
+\frac 1{2e_2}(\dot\omega_2^\mu)^2\ee
with
\be
\dot\omega_i^\mu=\dot x_i^\mu+i\theta_i\sigma^\mu \dot{\bar\theta}_i-i \dot\theta_i \sigma^\mu\bar\theta_i,\quad i+1,2\ee
 where $x_i^\mu,\theta_i$ are the space-time
coordinates and Grassmann variables of the two particles.

This Lagrangian is invariant under the two superconformal groups acting on the variables of each particle.

Let us write 
 the SUSY transformations 
\bea
&\delta\theta_i=\epsilon_i,~~~\delta\bar\theta_i =\bar\epsilon_i&\nn\\
&\delta x_i^\mu=i\theta_i\sigma^\mu\bar\epsilon_i-i\epsilon_i\sigma^\mu\bar\theta_i,~~~\delta e_i=0,~~~~i=1,2 &\label{eq:21}\eea

In order to introduce the superconformal interactions, following 
the bosonic case \cite{Casalbuoni:2014ofa}, let us define a 
space-time relative variable:
 \be  x_{12}^\mu=x_1^\mu-x_2^\mu-i\theta_1\sigma^\mu\bar\theta_2+i\theta_2\sigma^\mu\bar\theta_1\ee
 and the relative spinors
 \be
 \theta_{12}=\theta_1-\theta_2,~~~~\bar\theta_{12}=\bar\theta_1-\bar\theta_2.
 \ee
It is easily verified that $x_{12}^\mu$, $\theta_{12}$ and $\bar\theta_{12}$ are invariant under  the SUSY transformations (\ref{eq:21}) with $\epsilon_1=\epsilon_2=\epsilon$,   i.e., the diagonal supersymmetry.
The transformation properties of these variables under superinversion are complicated. Instead the quantity
\be\label{eq:51}
d_{12}^2=x_{12}^2+\theta_{12}^2\bar\theta_{12}^2\to \Omega(x_1)\Omega(x_2)d_{12}^2,
\ee
is invariant up to a superconformal factor,
or, using (\ref{eq:17b})
\be
x_{12}^2+\theta_{12}^2\bar\theta_{12}^2\to \frac 1{(x_1^2+\theta_1^2\bar\theta_1^2)}(x_{12}^2+\theta_{12}^2\bar\theta_{12}^2)\frac 1{(x_2^2+\theta_2^2\bar\theta_2^2)}
.\label{eq:26}\ee
The equations (\ref{eq:16}) and (\ref{eq:26}) generalise the transformation properties of $\dot x^2$ and $(x_1-x_2)^2$ of the nonsupersymmetric case \cite{Casalbuoni:2014ofa}
\be
\dot x^2\to\frac 1 {x^4} \dot x^2,~~~~(x_1-x_2)^2\to \frac 1{x_1^2}(x_1-x_2)^2\frac 1{x_2^2}\ee
In other words, the conformal factor $1/x^2$ goes into the superconformal factor $\Omega(x)$ (\ref{eq:17b}).




We are now in the position of writing down a two superconformal particle interaction. 
A possible superconformal model for two interacting superparticles is given by the action

\be
S_2=\int d\tau L_2=\int d\tau\left(\frac 1{2e_1}\dot\omega_1^2+\frac 1{2e_2}\dot\omega_2^2+\frac{\alpha^2}4\frac{\sqrt{e_1e_2}}{d_{12}^2}\right)
\label{eq:31}\ee
where $d_{12}^2$ is given in eq.~(\ref{eq:51}).
The transformation properties  of the variables under dilatations are given by 
\be
x_i^\mu\to\lambda x_i^\mu,~~~\theta_i\to\lambda^{1/2}\theta_i,~~~e_i\to \lambda^2 e_i,~~~~i=1,2.\label{eq:32}\ee
The SUSY transformations are contained in   (\ref{eq:21}). Instead  under superinversions, eq.~(\ref{eq:4}), we have
\be
\dot\omega_i^2\to\frac{\dot\omega_i^2}{(x^2+\theta_i^2\bar\theta_i^2)^2}\equiv\Omega^2(x_i){\dot\omega_i^2},~~~~i=1,2,
\label{eq:33}\ee
and 
 for the einbeins:
\be
e_i\to \Omega^2(x_i) {e_i},~~~~i=1,2.\label{eq:34}\ee
The action $S_2$ is superconformal invariant.

 In order to obtain the action in terms of superconfiguration variables $x^\mu{}_i, \theta_i$ we compute the equation of motion of the einbein variables
$e_i$

\be
\frac{\de L_2}{\de e_1}=-\frac{\dot\omega_1^2}{2e_1^2}+\frac{\alpha^2}8\sqrt{\frac{e_2}{e_1}}\frac 1{d_{12}^2}=0,\nn\ee
\be\frac{\de L_2}{\de e_2}=-\frac{\dot\omega_2^2}{2e_2^2}+\frac{\alpha^2}8\sqrt{\frac{e_1}{e_2}}\frac 1{d_{12}^2}=0.\label{eq:36} \ee
Solving these equations in $e_1$ and $e_2$ (the choice of the minus signs is for later convenience)
\be
\frac 1{e_1}=-\frac\alpha{2\dot\omega_1^2}\left(\frac{\omega_1^2\omega_2^2}{d_{12}^4}\right)^{1/4},~~~\frac 1{e_2}=-\frac\alpha{2\dot\omega_2^2}\left(\frac{\omega_1^2\omega_2^2}{d_{12}^4}\right)^{1/4}\ee
and substituting inside eq.~(\ref{eq:31}) we obtain the
action in superconfiguration space
\be
S_2=-\alpha\int d\tau\left(\frac{\dot\omega_1^2\dot\omega_2^2}{d_{12}^4}\right)^{1/4}.
\label{eq:89}\ee

 Notice that this action can be obtained from the bosonic configuration action of \cite{Casalbuoni:2014ofa}
\be
S_2=-\alpha\int d\tau\left(\frac{\dot{x}_1^2\dot{x}_2^2}
{(x_1-x_2)^4}\right)^{1/4},
\label{eq:89a}\ee
by the supersymmetric substitution
\be
\dot{x}_i^\mu\rightarrow\dot{\omega}_i^\mu, \quad
(\dot{x}_1-\dot{x}_2)^2\rightarrow d_{12}{}^2.
\ee

\subsection{Constraint analysis}
In order to do the constraint analysis here we consider the superconfiguration Lagrangian (\ref{eq:89}).
The conjugated momenta to $\dot{x}_i$ are given by
\be
p_1^\mu=\f {\de L}{\de\dot{x}_{1\mu}}= \f1 2\left(\f{\dot\omega_2^2}{ d_{12}^4}\right)^{1/4}\f {\dot\omega_1^\mu}{(\dot\omega_1^2)^{3/4}},\quad
   p_2^\mu=\f {\de L}{\de\dot{x}_{2\mu}}= \f1 2\left(\f{\dot\omega_1^2}{ d_{12}^4}\right)^{1/4}\f {\dot\omega_2^\mu}{(\dot\omega_2^2)^{3/4}},\quad
\ee
from which we obtain the primary constraint
\be
\phi=p_1^2p_2^2 -\frac{\alpha^4}{16 d_{12}^4}=0.
\ee
The fermionic momenta are given by
\be
\Pi_i=\frac {\de L_2}{\de\dot\th_i}=-i p_{i\mu}\s^\mu\bar\th_i,
\quad
\bar \Pi_i=\frac {\de L_2}{\de\dot{\bar\th}_i}=-i \th_ip_{i\mu}\s^\mu,~~~~i=1,2,
\ee
which imply four primary fermionic constraints
\be
D_i=\Pi_i+ip_{i\mu}\s^\mu\bar\th_i=0,\quad \bar D_i=\bar \Pi_i+i\th_ip_{i\mu}\s^\mu=0,~~~~i=1,2.
\label{di}
\ee
The Poisson brackets of the constraints (\ref{di}) are
\be
\{D_i,D_j\}=\{\bar D_i,\bar D_j\}=0,~~~~i,j=1,2
\ee
and
\be
\{D_i,\bar D_j\}=-2 i\delta_{ij}p_i\cdot \s,~~~~i,j=1,2.
\ee
Furthermore we have
\be
det\vert \{D_i,\bar D_j\}\vert =16p_1^2p_2^2=\frac {\alpha^4}{x_{12}^2+\th_{12}^2{\bar \th}_{12}^2},~~~~i,j=1,2.
\label{eq:68}
\ee
The determinant of the matrix of the fermionic constraint Poisson brackets given in eq.~(\ref{eq:68}) is different from zero unless one considers $r_{12}\to\infty$
and therefore the set of constraints $D_i,\bar D_j$ is second class.

Notice that  the presence of the interaction term modifies the structure of the constraint algebra with respect to the case of the free superconformal particle, giving a sort of  effective mass to the two superconformal particles: all fermionic constraints $D_i,\bar D_{ j}$ becomes second class as for the massive superparticle \cite{Casalbuoni:1976tz}.

The Dirac Hamiltonian is given by
\be
H_D=\lambda \phi +\sum_{i=1,2}\mu_i D_i+\sum_{i=1,2}\bar\mu_i \bar D_i,
\ee
and  the stability of the primary constraints gives
\be
0=\{\phi,H_D\}=\sum_{i=1,2}\mu_i \{\phi,D_i\}+\sum_{i=1,2}\bar\mu_i \{\phi, \bar D_i\},
\label{eq:57}
\ee
\be
0=\{D_i,H_D\}=\lambda \{D_i,\phi \}-\bar \mu_i\{D_i,\bar D_i\},\quad i=1,2,
\label{eq:58}
\ee
\be
0=\{\bar D_i,H_D\}=\lambda \{\bar  D_i,\phi \}- \mu_i\{\bar  D_i, D_i\}\quad i=1,2.
\label{eq:59}
\ee
By solving eqs.~(\ref{eq:58}) and (\ref{eq:59}) for $\mu_i$ and $\bar\mu_i$ and substituting in eq.~(\ref{eq:57}) we obtain  { the first class Dirac Hamiltonian
\be
H_D=\lambda \left [\phi +\sum_{i=1,2}\{\bar D_i, \phi\}\{D_i,\bar D_i\}^{-1}D_i +\sum_{i=1,2}\{D_i,\phi\}\{D_i,\bar D_i\}^{-1}\bar D_i\right ].
\ee
In conclusion we have a first class constraint
\be
\tilde \phi=\phi +\sum_{i=1,2}\{\bar D_i,\phi \}\{D_i,\bar D_i\}^{-1}D_i +\sum_{i=1,2}\{ D_i,\phi\}\{D_i,\bar D_i\}^{-1}\bar D_i 
\ee
and four second class constraints
\be
D_{i\alpha},\quad \bar D_{i\dot \alpha},~~~~i,j=1,2.
\ee
Since in this case there is only one primary constraint that generates worldline diffeomorphism, there is no kappa symmetry.

\subsection{Quantization}
Quantization can be performed \`a la Gupta Bleuler by requiring  the following operatorial conditions on the "ket" vectors
\be
\tilde \phi |\Phi>=0,\quad \bar D_{i\dot \alpha}|\Phi>=0
\label{eq:87}
\ee
and  the following ones on the "bra":
\be<\Phi|D_{i\alpha}=0.
\label{eq:64}
\ee
Note that the solution to the second one of Eqs.~(\ref{eq:87})  implies that the bilocal field $\Phi(x_i,\th_i,\bar\th_i)=<x_i,\th_i,\bar\th_i|\Phi>$ is a bilocal chiral superfield
\be  
\Phi(x_i,\th_i,\bar\th_i)=\Phi(\th_i,y_i),
\ee
where
\be
y_i^\mu=x_i^\mu+i \th_i\s^\mu\bar\th_i,\quad i=1,2.
\ee

By using eq.~(\ref{eq:87}) we have
\be
\tilde \phi |\Phi>=\left [\phi +\sum_{i=1,2}\{\bar D_i,\Phi\}\{D_i,\bar D_i\}^{-1} D_i\right ]|\Phi>=0.
\label{eq:114}
\ee
Note that $\tilde \phi |\Phi>$ is also a chiral superfield. Indeed
\be
[\bar D_{i\a},\tilde \phi]=0
\ee
implies
\be
\bar D_{i\a}\tilde \phi|\Phi>=0.
\ee

Chiral bilocal superfield can be expanded as
\bea
\Phi(\th_i,y_i)&=&\phi(y_i)+\th_i^\a\psi_\a^i(y_i)+\th_i^\a\th_j^\beta \Big [\epsilon^{ij}F_{\mu\nu}(y_i)\s^{\mu\nu}_{\a\beta}\\
\nn
&+&\epsilon_{\a\beta}C^{ij}(y_i)\Big ]+\th_2^2\th_1^\a\chi_{1\a}(y_i)+\th_1^2\th_2^\a\chi_{2\a}(y_i)+\th_1^2\th^2_2F(y_i)\eea
and contains five scalars $\phi, C^{ij},F$,  a  3-component antisymmetric tensor $F_{\mu\nu} $ and eight  fermionic fields $\psi_\a^i, \chi_i^\a$.

Wave equations for the component fields can be evaluated  by expanding eq.~(\ref{eq:114}) in series of  Grassmann variables $\th_i$. For the scalar field $\phi$ one recovers the field equation of the purely bosonic case \cite{Casalbuoni:2014ofa}, while for the fermionic and the other bosonic fields additional terms are present. This analysis is beyond the aim of the present paper and deserves further studies.

\section{Nearest-neighbour interactions and continuum limit}

 In this Section we generalise the model 
by considering  a system of superconformal particles  in which each particle interacts with its nearest neighbours. In other words we will consider the $N+1$ particles as an ordered set labelled by an index $i$ running from 1 to $N+1$ on a one-dimensional lattice with a lattice spacing denoted by $a$.
  
We assume the following action, containing only two-body interactions of the type that we have already proposed in Section \ref{twosuperconformal},
\be
S=\int d\tau \left[\sum_{i=1}^{N+1}\frac{\dot \omega_i^2}{2e_i} +\frac {\alpha^2}4 \sum_{i=1}^N \frac{\sqrt{e_i e_{i+1}}}{d_{i,i+1}^2}\right]\label{eq:213}\ee
with
\be d_{i,i+1}^2=x_{i,i+1}^2+\theta_{i,i+1}^2\bar\theta_{i,i+1}^2\ee
and
  \be  x_{i,i+1}^\mu=x_i^\mu-x_{i+1}^\mu-i\theta_i\sigma^\mu\bar\theta_{i+1}+i\theta_{i+1}\sigma^\mu\bar\theta_i,\quad
 \theta_{i,i+1}=\theta_i-\theta_{i+1},~~~~\bar\theta_{i,i+1}=\bar\theta_i-\bar\theta_{i+1}.
 \ee

 Instead of considering a linear lattice one could identify the two ends $x_1=x_{N+1}$, and close the lattice to a circle. 
  Let us notice that the physical dimensions of the various quantities appearing in this Lagrangian are
 $
 [x]=[\tau]=[e]=\ell,~~[\theta]=[\bar\theta]=\ell^{1/2},~~[\alpha]=\ell^0
 $.

Here, we will not discuss this action but rather its continuum limit.
 To this end, let us define a variable $\sigma$ to identify the lattice points
 \be \sigma_i= ia,~~~i=1,\cdots, N+1\ee
  
 In the continuum limit we have
 \bea
 & \f 1 {\dd a} x_{i,i+1}^\mu \to \dd{ -\left[\frac{\de x^\mu}{\de \sigma}+i\theta\sigma^\mu\frac{\de\bar\theta}{\de\sigma}-i\frac{\de\theta}{\de\sigma}\sigma^\mu\bar\theta\right]}
 \equiv -\omega^{\mu\prime}, \quad
 & \f 1 {\dd a}  \theta_{i,i+1}\to -\frac{\dd\de\theta}{\dd \de\sigma}
 \equiv \theta{'}
 \eea  
 and
 analogously for $\bar\theta_{i,i+1}$.
 Notice that $\omega'$ transforms under superconformal inversion exactly as $\dot\omega$, that is
 \be
\omega^{'2}\to\frac{\omega^{'2}}{(x^2+\theta^2\bar\theta^2)^2}\equiv\Omega^2(x)\omega^{'2}.\ee

   Furthermore, the sum must be transformed as follows
  \be
  \sum_i \to \frac 1 a \int d\sigma .\ee
  The expression (\ref{eq:213}) becomes (assuming $a=\pi/(N+1)$ or $\sigma$ to vary in the range $(0,\pi)$)
  \be
  S\to -\int  d\tau\int_0^\pi d\sigma\left[ \frac 1 a\frac {\dot \omega^2(\sigma,\tau)}{2e(\sigma,\tau)}+\frac 1{a^3}\frac{\alpha^2}4\frac{e(\sigma,\tau)}{\omega^{'2}(\sigma,\tau)}\right].\ee
     In order to eliminate the divergence we  redefine the einbein field   $e(\sigma,\tau)$  and the coupling $\a$ as
        \be
   {\tilde e}= a e,\quad   \frac{\tilde\alpha^2}2=\frac 1{a^4}\frac{\alpha^2}4 \ee
 where the factor 1/2 has been chosen for later convenience. Then, by denominating $e$ and $\a$  as before, we obtain the action in the continuum limit:
\be S=\int d\tau\int_0^\pi d\sigma\left[ \frac 1 2 \frac {\dot\omega^2(\sigma,\tau)}{e(\sigma,\tau)}+\frac{\alpha^2}2\frac{e(\sigma,\tau)}{\omega^{'2}(\sigma,\tau)}\right].\label{eq:222}\ee
  By  varying the action with respect to $e(\sigma,\tau)$ we get
   \be
   \frac 12 \frac{\dot \omega^2}{e^2}= \frac{\alpha^2}2 \frac 1{\omega^{'2}}\ee
  or
   \be
   e=\frac 1 \a \sqrt{\dot \omega^2 \omega^{'2}}.\ee
   By substituting the expression of the einbein inside eq.~(\ref{eq:222}) we get 
   \be\label{ngconformal}
   S=-\alpha\int d\tau\int_0^\pi d\sigma\sqrt{\frac{\dot \omega^2}{\omega^{'2}}}.\ee
   Notice that the action is trivially  conformal invariant, since $\dot \omega^2$ and $\omega^{'2}$  transform in the same way under inversion. It is also invariant under diffeomorphism in $\tau$ but not in $\sigma$.
 In this paper we do not perform the constraint analysis and their physical consequences. 

\section{Outlook}
For future investigations it would be interesting to analyse several aspects that we did not consider in this paper, starting, for example by a study of  the equations of motion for the components of the bilocal chiral superfield and their solutions. It would  also be interesting to compare the results of the predictions of the  weak and the reduced space quantization. As already noted in the paper, another subject which deserves further work is the analysis of the constraints and their physical consequences in the continuum limit of the model. 
The study of the Killing equation would be interesting to find if by any chance the model contains some accidental symmetry. Finally, future investigations will be devoted to the Carroll and non-relativistic limits of the model.

\acknowledgements
We would like to thank Paul Townsend and Kiyoshi Kamimura for useful comments.
One of us (JG) would like to thank the Galileo Galilei Institute for Theoretical
Physics and the INFN for partial support during the completion of this
work. The research of JG was supported in part by PID2019-105614GB-C21 and by the State
Agency for Research of the Spanish Ministry of Science and Innovation through the Unit of
Excellence Maria de Maeztu 2020-2023 award to the Institute of Cosmos Sciences (CEX2019-
000918-M).



\end{document}